\documentclass[titlepage,12pt]{article}
\usepackage{amssymb,amsmath,color,graphics,amscd,epsf,indentfirst,amsfonts}
\usepackage{epsfig}
\usepackage[dvipdfm,hypertex]{hyperref}
\usepackage{nicefrac}
\usepackage{scalefnt}
\usepackage[titles]{tocloft}
\usepackage{sectsty}
\allsectionsfont{\bf \scalefont{.7} \selectfont}
\subsectionfont{\bf \scalefont{.85} \it \selectfont}
\subsubsectionfont{\bf \scalefont{1} \it \selectfont}
\usepackage[T1]{fontenc}
\usepackage{lmodern}
\usepackage{bibspacing}

\def\blfootnote{\xdef\@thefnmark{}\@footnotetext}

\long\def\symbolfootnote[#1]#2{\begingroup%
\def\thefootnote{\fnsymbol{footnote}}\footnote[#1]{#2}\endgroup}

\makeatletter
\renewcommand{\@dotsep}{4.5}
\makeatother

\def\be{\begin{equation}}
\def\ee{\end{equation}}

\makeatletter
\def\@seccntformat#1{\csname the#1\endcsname.\quad}
\makeatother

\def\clock{{\count0=\time
           \divide\count0 60
           \ifnum\count0<10 0\fi\the\count0
           \multiply\count0 -60 \advance\count0 \time
           :\ifnum\count0<10 0\fi \the\count0
         }}
\newcommand{\timestamp}{{\small\vbox{\hbox{\tt\jobname.tex}
\hbox{\the\day/\the\month/\the\year, \clock}}}}

\def\MM{{\cal M}}
\def\NN{{\cal N}}

\newcommand{\beq}{\begin{equation}}
\newcommand{\eeq}{\end{equation}}
\newcommand{\ba}{\begin{array}}
\newcommand{\ea}{\end{array}}
\newcommand{\bea}{\begin{eqnarray}}
\newcommand{\eea}{\end{eqnarray}}

\newcommand{\Z}{\mathbb{Z}}

\newcommand{\R}{\mathbb{R}}

\newcommand{\tr}{\mathop{{\rm Tr}}}

\numberwithin{equation}{section}

\begin{document}

\begin{titlepage}
\begin{flushright}
CPHT-RR 057.0808\\
\end{flushright}
\vskip 2.8cm
\begin{center}
\font\titlerm=cmr10 scaled\magstep4
    \font\titlei=cmmi10 scaled\magstep4
    \font\titleis=cmmi7 scaled\magstep4
    \centerline{\titlerm
      Seiberg Duality in Chern-Simons Theories
      \vspace{0.4cm}}
    \centerline{\titlerm
      with Fundamental and Adjoint Matter}   
\vskip 1.5cm
{\it Vasilis Niarchos}\\
\vskip 0.7cm
\medskip
{Centre de Physique Th\'eorique, \'Ecole Polytechnique,
91128 Palaiseau, France}\\
{Unit\'e mixte de Recherche 7644, CNRS}\\

\end{center}
\vskip .4in
\centerline{\bf Abstract}

\baselineskip 20pt
%

\vskip .5cm \noindent
We explore the dynamics of three-dimensional Chern-Simons 
gauge theories with $\NN=2$ supersymmetry and matter in the
fundamental and adjoint representations of the gauge group.
Realizing the gauge theories of interest in a setup of threebranes
and fivebranes in type IIB string theory we argue for a Seiberg
duality that relates Chern-Simons theories with non-trivial 
superpotentials.

\vfill
\noindent
August 2008
\end{titlepage}\vfill\eject

\setcounter{equation}{0}

\pagestyle{empty}
\small
\normalsize
\pagestyle{plain}
\setcounter{page}{1}

\section{Introduction}
\label{sec:intro}

There has been a recent resurgence of interest in Chern-Simons (CS) theories 
with varying amounts of supersymmetry. Chern-Simons theory without matter
is a topological three dimensional field theory. It ceases to be topological,
however, when it is coupled to matter. Then it exhibits non-trivial dynamics which 
are interesting for several reasons.

Conformal Chern-Simons theories with $\NN=8$ supersymmetry are expected to
describe the low energy worldvolume dynamics of M2-branes in M-theory. Ref.\ 
\cite{Schwarz:2004yj} explored various Chern-Simons theories with this purpose,
but did not find one with $\NN=8$ supersymmetry. Theories with this amount of
supersymmetry were constructed in \cite{Bagger:2006sk,Bagger:2007jr,Bagger:2007vi,
Gustavsson:2007vu,Gustavsson:2008dy} and were related to M2-brane dynamics
in \cite{Lambert:2008et,Distler:2008mk}. Theories with $\NN=6$ and $\NN=5$ 
supersymmetry were recently discussed in \cite{Aharony:2008ug,Aharony:2008gk}. 
A non-supersymmetric variant was considered in \cite{Armoni:2008kr}.

Via the AdS/CFT correspondence conformal Chern-Simons theories map to a 
class of four dimensional string/M theory backgrounds with negative cosmological 
constant. By studying them we may hope to learn more about string/M theory in these 
backgrounds. 

Chern-Simons theories also arise in interesting condensed matter systems (see $e.g.$ 
\cite{Polychronakos:1989cd,Polychronakos:1990xq,Iengo:1992du,wen:1995,
Cooper:1997uz}). These include systems that exhibit quantum Hall effects or 
superconductivity. Supersymmetric Chern-Simons-matter theories are interesting 
in this respect as solvable toy models. 

In this note we will focus on Chern-Simons-matter theories with $\NN=2$ 
supersymmetry ($i.e.$ four real supercharges). These theories are characterized by 
a gauge group $G$, the Chern-Simons level $k$ and the matter representations
$R_i$ \cite{Gaiotto:2007qi}. For non-abelian gauge groups the level $k$ is 
quantized. We will restrict to situations where the gauge group $G$ is the 
unitary group $U(N)$.

By adding superpotential interactions among the matter superfields one can 
break the conformal invariance and generate non-trivial renormalization
group (RG) flows. In these situations, one would like to be able to determine 
the infrared (IR) behavior of the theory. In four dimensional gauge theories
with four supercharges ($i.e.$ $\NN=1$ supersymmetry in 4d terms), there 
has been considerable progress in understanding such flows. Important tools
in this progress are the NSVZ $\beta$-function formula \cite{Novikov:1983uc},
Seiberg duality \cite{Seiberg:1994pq}, $a$-maximization \cite{Intriligator:2003jj,
Kutasov:2003ux}, $etc.$ Similar progress in three dimensions would be desirable.

The recent reference \cite{Giveon:2008zn} has taken a first step in this direction
by proposing a Seiberg duality for $\NN=2$ Chern-Simons-matter theories with
gauge group $U(N_c)$ and $N_f$ pairs of chiral multiplets $Q^i, \tilde Q_i$
($i=1,2,\cdots,N_f$). The superfields $Q^i$ are in the fundamental representation 
of the gauge group and $\tilde Q_i$ in the anti-fundamental. A close cousin of this 
theory in four dimensions is $\NN=1$ SQCD. It is our purpose here to take a further 
step along these lines by studying $\NN=2$ Chern-Simons-matter theories with
additional chiral multiplets in the adjoint representation. The analog of these theories 
in four dimensions is $\NN=1$ SQCD theories with adjoint chiral superfields 
\cite{Kutasov:1995ve,Kutasov:1995np,Kutasov:1995ss,Intriligator:2003mi}.
We will postulate a Seiberg duality for these theories and provide some
checks. We will focus mostly on the case of one adjoint chiral superfield.
An interesting subtlety of a similar exercise with two adjoint chiral
superfields will also be mentioned.

Section \ref{elec} formulates the theory of interest. We will find it convenient
to phrase our statements in the language of brane configurations in type IIB
string theory. Section \ref{mag} argues for a Seiberg duality and provides 
some checks. We conclude in section \ref{end} with a summary of the main
lessons and a list of interesting open problems.

\section{The electric theory}
\label{elec}

We will realize the gauge theories of interest as low energy effective
field theories residing in a configuration of threebranes and fivebranes
in type IIB string theory in $\R^{9,1}$. This will provide a quick and
intuitive access to many classical and quantum aspects of CS dynamics,
which can be formulated, of course, independently in field theory language.

\begin{figure}[t!]
\centering
\includegraphics[width=10.2cm, height=7.2cm]{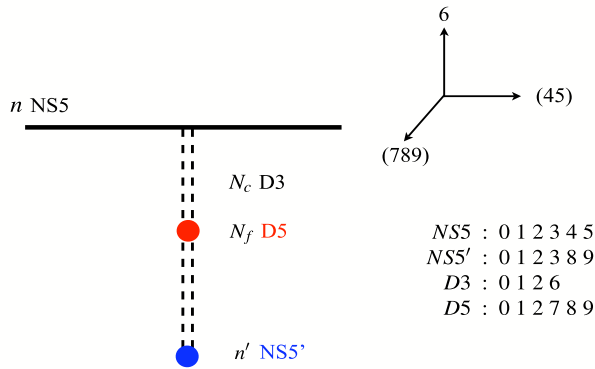}
\vspace{-1.5cm}
\caption{\it A configuration of D3, D5 and NS5-branes that realizes
$\NN=2$ $U(N_c)$ SQCD in three dimensions with two extra adjoint 
chiral superfields.}
\label{SQCD}
\end{figure}

An instructive precursor of the configurations we want to consider appears in 
fig.\ \ref{SQCD}.
This configuration preserves four supercharges, $i.e.$ it exhibits $\NN=2$
supersymmetry in the three directions $(x^0,x^1,x^2)$ common to all the
branes.\footnote{With both bunches of the NS5-branes parallel along (012345), 
instead of orthogonal, we would obtain $\NN=4$ supersymmetry. That configuration,
with $n=n'=1$, is the one analyzed in the original work \cite{Hanany:1996ie}.} 
The low energy description of this system is in terms of a $U(N_c)$
gauge theory that lives on the D3-branes which are suspended along the 
6-direction between the $n$ NS5-branes and the $n'$ NS5$'$-branes.
The matter content of this theory comprises of: $(a)$ an $\NN=2$ vector
multiplet $V$, $(b)$ $N_f$ pairs of $\NN=2$ chiral multiplets $Q^i$, $\tilde Q_i$
($i=1,2,\cdots,N_f$) in the fundamental and anti-fundamental representations of
the gauge group and $(c)$ two chiral supermultiplets $X$, $Y$ in the adjoint
representation. 

The vector multiplet $V$ arises from 3-3 strings on the D3-branes and includes
both the gauge field $A_\mu$ $(\mu=0,1,2)$ and a scalar $\sigma$. The scalar
parametrizes the position of the D3-branes in the $x^3$ direction. The $N_f$
pairs of chiral multiplets $Q$, $\tilde Q$ arise from the 3-5 strings stretching
between the D3 and the D5-branes. Finally, the chiral multiplets $X,Y$ arise from 
3-3 strings and their scalar components parametrize the position of the D3-branes
along the $(89)$ and $(45)$ directions respectively. 

The dynamics of the vector multiplet $V$ and chiral multiplets $Q,\tilde Q$
are those of $\NN=2$ SQCD in three dimensions. The extra adjoint chiral superfields
have a non-trivial superpotential which fixes the values of their scalar components.
We will not discuss further the dynamics of this system since it lies outside our
scope. Instead, we will now proceed to consider a closely related theory that 
arises from that of fig.\ \ref{SQCD} by a certain deformation.

Let us only note in passing that by compactifying the direction $x^3$ and T-dualizing 
the configuration of fig.\ \ref{SQCD} we obtain in type IIA string theory a configuration 
that realizes at low energies $\NN=1$ SQCD in four dimensions with two adjoint chiral 
superfields \cite{Elitzur:1997hc,Giveon:1998sr}. This theory has a polynomial 
superpotential of the form 
\beq
\label{elecaa}
W=\frac{s_0}{n+1}\tr X^{n+1}+\frac{s_0'}{n'+1}\tr Y^{n'+1}+\tr [X,Y]^2+
\tilde Q_i Y Q^i
~.
\eeq
In the brane configuration $s_0,s_0'$ should be thought of as large numbers.

\begin{figure}[t!]
\centering
\includegraphics[width=10.4cm, height=7.4cm]{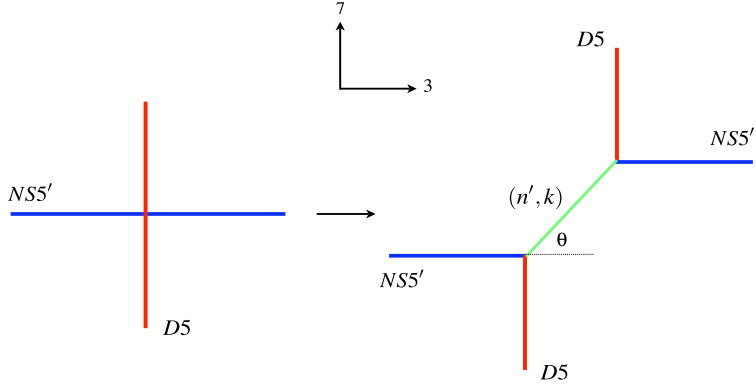}
\vspace{-1.5cm}
\caption{\it $k$ D5-branes recombining with $n'$ NS5-branes to give
an $(n',k)$ fivebrane bound state.}
\label{recomb}
\end{figure}

\begin{figure}[t!]
\centering
\includegraphics[width=10.2cm, height=7.2cm]{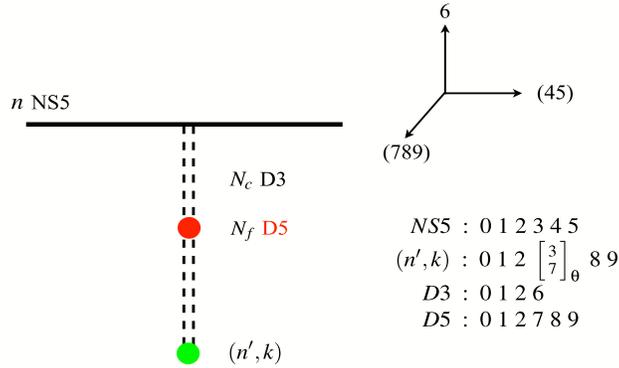}
\vspace{-1.5cm}
\caption{\it The configuration of branes after the reconnection of
$k$ D5-branes and $n'$ NS5$'$-branes. The notation $\left[ {3 \atop 7}\right]_\theta$
denotes the fact that the $(n',k)$ brane is rotated in the $(37)$ plane
with an angle $\theta$.}
\label{CSQCD}
\end{figure}

The deformation we want to consider is the following. Start with the configuration
in fig.\ \ref{SQCD} with $N_f+k$ D5-branes. Then, move $k$ of these D5-branes
along the $x^6$ direction until they intersect with the $n'$ NS5$'$-branes. At this point
we can locally reconnect the D5 and NS5$'$ branes as in fig.\ \ref{recomb} to obtain 
an $(n',k)$ fivebrane bound state. The resulting configuration will continue to preserve 
the same amount of supersymmetry provided that the $(n',k)$ fivebrane is rotated in the 
$(37)$ plane with a specific angle $\theta$. This angle is determined by the integers 
$n',k$ via the relation \cite{Aharony:1997ju}
\beq
\label{elecab}
\tan \theta=g_s \frac{k}{n'}
~.
\eeq
After the reconnection we send the NS5$'$ and D5-branes to infinity to be left with 
the configuration of fig.\ \ref{CSQCD} whose dynamics are described at low energies by 
the theory we are interested in.

The low energy theory on the D3-branes suspended between the NS5
and $(n',k)$ branes still has $\NN=2$ supersymmetry and the low energy 
degrees of freedom are again given by the vector multiplet $V$, the $N_f$
pairs of chiral multiplets $Q^i$, $\tilde Q_i$ and the adjoint chiral multiplets 
$X,Y$. However, the dynamics of these fields are now different. 

Most notably, the rotation of the $(n',k)$ fivebrane in the $(37)$ plane suggests 
that the scalar $\sigma$ has become massive. By supersymmetry the whole vector 
multiplet must become massive. In three dimensions a vector field can become
massive without spoiling gauge invariance by adding to the Lagrangian the
Chern-Simons interactions. Indeed, it was argued in \cite{Kitao:1998mf,Bergman:1999na}
that the gauge theory on D-branes ending on $(p,q)$ fivebrane bound states
includes the Chern-Simons Lagrangian at fractional level $\frac{q}{p}$.
A fractional level is acceptable for a $U(1)$ gauge theory, but it is in
conflict with the level quantization in non-abelian gauge theories. For
a recent discussion on this point see \cite{Gaiotto:2008ak}. In order to 
avoid this issue, in the rest of this note we will restrict attention to the 
special case where $n'=1$. In that case, our configurations realize
a Chern-Simons theory with integer level $k$.

The dynamics of the remaining multiplets are unaffected 
by the deformation. Hence, we end up at low energies with an $\NN=2$
Chern-Simons theory at level $k$ coupled to $N_f$ pairs of chiral multiplets 
$Q^i,\tilde Q_i$ and two adjoint chiral superfields $X,Y$. For reasons that will
become apparent momentarily the theory also possesses a superpotential of
the form \eqref{elecaa}. For $n'=1$ the superfield $Y$ is massive and
can be integrated out. Then we are left with a single adjoint superfield, the 
superfield $X$, that has the superpotential 
\beq
\label{elecac}
W=\frac{s_0}{n+1}\tr X^{n+1}
~.
\eeq
As a first check, notice that for $n=1$ the superfield $X$ is also massive and
can be integrated out. Then, as anticipated, we recover the theory of ref.\ 
\cite{Giveon:2008zn}.

In the brane configuration of fig.\ \ref{CSQCD} (for $n'=1$) there are 
$N_c-nN_f$ D3-branes that have to stretch between the $n$ NS5-branes 
and the $(1,k)$ bound state. According to the $s$-rule of ref.\ \cite{Hanany:1996ie}
supersymmetry is preserved if the maximum of these branes is $nk$, hence the 
constraint
\beq
\label{elecad}
nN_f+nk-N_c \geq 0
~.
\eeq
In our $\NN=2$ CS theory this is a necessary property for the existence of a 
supersymmetric vacuum. It is worth comparing this condition to a corresponding 
condition for stability in the four dimensional $\NN=1$ SQCD with a single adjoint chiral
superfield. In that case the condition is $nN_f-N_c\geq 0$ \cite{Kutasov:1995np}.

The superpotential \eqref{elecac} can be deduced from the brane moduli
space in the following way. Displace the $n$ NS5-branes in the (89) plane
and place them at $n$ different points $a_j=x_j^8+ix_j^9$, $j=1,2,\cdots,n$.
Then, the $N_c$ D3-branes will also break up into $n$ groups of $r_1$ D3-branes
ending on the $a_1$ positioned NS5-brane, $r_2$ D3-branes ending on the 
$a_2$ positioned NS5-brane $etc.$ with 
\beq
\label{elecae}
\sum_{i=1}^n r_i=N_c
~.
\eeq
From the D3-brane point of view $a_i$ are the expectation values of the 
diagonal matrix elements of the complex scalar in the superfield $X$.
In order to account for these vacua in gauge theory a polynomial superpotential
is needed of the form
\beq
\label{elecaf}
W(X)=\sum_{i=0}^n \frac{s_j}{n+1-i}X^{n+1-i}
~.
\eeq
For generic coefficients $\{s_j\}$ the superpotential has $n$ distinct minima
$\{ a_j\}$ related to $\{s_j \}$ via the relation
\beq
\label{elecag}
W'(x)=\sum_{i=0}^n s_i x^{n-j}=s_0 \prod_{i=1}^n(x-a_i)
~.
\eeq
In the gauge theory picture the integers $(r_1,\cdots,r_n)$ label the number of
the eigenvalues of the $N_c\times N_c$ matrix $X$ residing in the $i$th minimum
(for $r_i$) of the potential $V=|W'(x)|^2$. When all the expectation values $a_j$ 
are distinct the adjoint field is massive and the gauge group is Higgsed
\beq
\label{elecai}
U(N_c)\to U(r_1)\times U(r_2)\times \cdots \times U(r_n)
~.
\eeq
In this vacuum we get $n$ decoupled copies of the $\NN=2$ CS theories
with fundamentals that were considered in \cite{Giveon:2008zn}.

The superpotential \eqref{elecac} is a classically relevant interaction 
for $n=1,2$. For $n=1$ we recover in the IR the theory of \cite{Giveon:2008zn}.
For $n=2$ we are driven towards a different IR theory whose precise properties are
unknown. For $n=3$ the interaction is classically marginal. It has been argued in 
\cite{Gaiotto:2007qi} that this interaction drives the theory towards an interacting IR 
fixed point. Finally, for $n\geq 4$ the interaction is classically irrelevant. In analogy
to the four dimensional case of $\NN=1$ adjoint-SQCD, we would like to propose that 
these interactions are in fact `{\it dangerously irrelevant}'. At large enough coupling, 
the corresponding operators develop large anomalous dimensions in the theory without 
the superpotential interaction and become relevant, hence they can affect the IR physics 
in a non-trivial manner when added to the Lagrangian. Unfortunately, the necessary 
technology is currently lacking to verify this statement explicitly.

The global symmetry of the theory is 
\beq
\label{elecaj}
SU(N_f)_v\times SU(N_f)_a \times U(1)_a \times U(1)_R
~.
\eeq
The first three of these symmetries become obvious in the brane setup 
by moving the $N_f$ D5-branes along $x^6$ on top of the $(1,k)$ fivebrane
and performing separate $U(N_f)$ transformation on the portions of the 
D5s with $x^7>0$ and $x^7<0$. The last one is an R-symmetry. The
theory has two R-symmetries, but only under one of them is the superpotential
\eqref{elecac} invariant. In the brane setup these symmetries are related to
the geometric rotation symmetries $U(1)_{45}$, $U(1)_{89}$ along the 
(45) and (89) planes respectively.

Other deformations of the field theory involving the quark superfields $Q^i$, 
$\tilde Q_i$ are also easy to see in the brane picture. For example, moving 
the D5-branes in the (45) plane corresponds in field theory to turning on
complex masses for $Q,\tilde Q$ via the superpotential interaction 
$W=m_i\tilde Q_i Q^i$. Moving the D5s in the $x^3$ direction corresponds
to turning on real masses with opposite signs to $Q$, $\tilde Q$.

The field theory has a large moduli space $\MM$ parametrized by the expectation 
values of the scalar components of the quark superfields $Q^i$, $\tilde Q_i$.
This space arises in the brane construction by separating the $N_f$ D5-branes
in the $x^6$ direction and then splitting the D3-branes on them \cite{Elitzur:1997hc,
Giveon:1998sr}. The complex dimension of the moduli space is 
\beq
\label{elecal}
\dim \MM=\Bigg \{ {{nN_f^2~, ~~~ \hspace{4cm} N_f \leq m} \atop 
{2N_fN_c-nm^2-p(2m+1)~, ~~~ N_f> m}}
~
\eeq
where we decomposed the number of colors $N_c$ as
\beq
\label{elecak}
N_c=nm+p~, ~ ~ m,~p\in \Z_{\geq 0}~, ~ ~0\leq p<n
~.
\eeq

\section{The magnetic theory}
\label{mag}

\begin{figure}[t!]
\centering
\includegraphics[width=10.2cm, height=7.2cm]{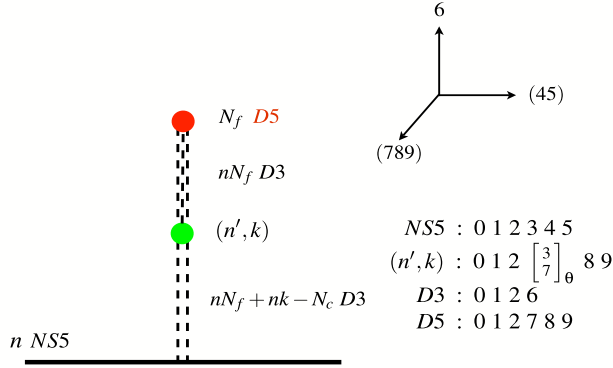}
\vspace{-1.5cm}
\caption{\it The magnetic configuration of branes for general $n,n'$.}
\label{CSQCDmag}
\end{figure}

By moving the D5-branes and the $(1,k)$ fivebrane past the $n$ NS5-branes
along the $x^6$ direction we obtain, as in \cite{Elitzur:1997hc,Giveon:1998sr}, 
the configuration that appears in fig.\ \ref{CSQCDmag}. When a $(p,q)$5-brane 
passes through a $(p',q')$5-brane $|qp'-pq'|$ D3-branes are created 
\cite{Kitao:1998mf}. Hence, in fig.\ \ref{CSQCDmag} (for $n'=1$) $nN_f$ D3s are 
stretched between the D5s and the $(1,k)$ fivebrane and $nN_f+nk-N_c$ D3s 
are stretched between the $(1,k)$ fivebrane and the $n$ NS5s.

Assuming that the infrared dynamics are not affected by this process we
end up with a gauge theory which is Seiberg dual to the original. The dual theory
lives on the D3-branes stretching between the $(1,k)$ fivebrane and the $n$
NS5-branes. It is $\NN=2$ CS at level $k$ with gauge group $U(nN_f+nk-N_c)$, 
$N_f$ pairs of chiral multiplets $q_i,\tilde q^i$, an adjoint chiral superfield $Y$ and 
$n$ magnetic mesons $M_i$ $(i=1,\cdots,n)$, each of which is an $N_f\times N_f$
matrix. The magnetic mesons arise from 3-3 strings residing on the $nN_f$ D3-branes 
stretching between the D5s and the $(1,k)$ fivebrane.\footnote{A naive counting of the
Chan-Paton indices for 3-3 strings appears to give $nN_f \times nN_f$ massless fields
at the origin of moduli space. It has been pointed out, however, in a related context in 
\cite{Elitzur:1997hc} that this counting is misleading (see sec.\ 7.3 of \cite{Elitzur:1997hc}).} 
As in \cite{Elitzur:1997hc}, a superpotential of the form
\beq
\label{magaa}
W=-\frac{s_0}{n+1}\tr Y^{n+1}+\sum_{i=1}^n M_i \tilde q Y^{n-i}q
\eeq
is anticipated. We notice that the rank of the dual gauge group is 
$\tilde N_c=nN_f+nk-N_c$. The positivity of this rank is equivalent to 
the condition \eqref{elecad} for the existence of a supersymmetric vacuum in the
electric theory.

A potential concern for the validity of Seiberg duality in this system stems
from the fact that in the above transformation there is a singularity when
the NS5-branes meet with the $(1,k)$  fivebrane during their motion along $x^6$.
In the case of NS5/NS5$'$ configurations as in \cite{Elitzur:1997hc}, this 
singularity can be avoided by separating the NS5 and NS5$'$-branes along
their common transverse direction $x^7$. This is not possible, however, in our
configuration, since the $(1,k)$ fivebrane is rotated with some non-zero angle 
along the (37) plane. With this point noted, let us accept as a working assumption 
that the CS theories of this and the previous section are indeed dual to each other 
and see if we can make any checks of this tentative duality.

Repeating the arguments of ref.\ \cite{Giveon:2008zn} one can show that
the proposed duality is consistent with the structure of the moduli space and
deformations. The magnetic configuration has the same global symmetries 
\eqref{elecaj} as the electric configuration. This can be seen directly in the brane
setup as in the previous section. 

The moduli space of the magnetic configuration arises by separating the 
D5-branes along the $x^6$ direction and then splitting the $nN_f$ D3-branes 
stretched between them and the $(1,k)$ fivebrane consistently with the geometry 
\cite{Giveon:1998sr}. Counting the dimension of the resulting moduli space one
recovers the expressions of the electric case \eqref{elecal}. As in ref.\ 
\cite{Giveon:2008zn} it is important to notice in this exercise that when $N_f>m$
the $n(N_f+k)-N_c$ D3-branes stretching between the $(1,k)$ fivebrane
and the $n$ NS5-branes are more than $nk$ contrary to the $s$-rule. Then, 
to preserve supersymmetry one has to keep $nN_f-N_c$ flavor D3-branes at 
the origin. With this restriction one recovers the second expression of eq.\ \eqref{elecal}.
The agreement between the moduli spaces of the electric and magnetic theories
can also be deduced easily in the case of the general deformation \eqref{elecaf} 
and the associated Higgsing \eqref{elecai}. There is a corresponding deformation 
of the magnetic theory in this case with a superpotential 
\beq
\label{magab}
W_{magn}=-\sum_{i=0}^n \frac{\bar s_i}{n+1-i}\tr Y^{n+1-i}
+\sum_{i=1}^n \bar{\bar s}_i M_i\tilde q Y^{n-i}q
\eeq
where $\bar s_i=\bar s_i(\{s_j \})$, $\bar{\bar s}_i=\bar{\bar s}_i(\{s_j \})$ are functions
of $s_j$ whose precise form can be deduced with the methods of \cite{Kutasov:1995ss}.
The equality of the dimensions of the moduli spaces of the electric
and magnetic descriptions for each copy of the decoupled $\NN=2$ CS 
theories was checked in \cite{Giveon:2008zn}.

Several deformations of the electric theory can be matched directly to the
magnetic theory precisely as in \cite{Giveon:2008zn}. Since the analysis
presented there can be repeated here {\it mutatis mutandis} we will not be
explicit. As an example, we note that deforming
the electric theory by the superpotential $W=m_1 \tilde Q_1 Q^1$ corresponds
in fig.\ \ref{CSQCD} to separating one of the $N_f$ D5-branes in the (45) plane
from the D3-branes. In the magnetic description this deformation requires
one of the D3-branes connected to the D5-branes to combine with one of
the $nN_f+nk-N_c$ D3s thus reducing the gauge group by one.

\section{Closing remarks}
\label{end}

In this note we considered the possibility of Seiberg duality in Chern-Simons
theories with $\NN=2$ supersymmetry and matter in the fundamental and
adjoint representations. Generalizing the arguments of \cite{Giveon:2008zn}
to include a field in the adjoint representation we found evidence for a duality 
between $U(N_c)$ and $U(nN_f+nk-N_c)$ CS theories both at integer level
$k$. $N_f$ is the number of flavor chiral multiplets $Q^i,\tilde Q_i$. The
$U(1)$ part of the gauge groups is interacting and important in these theories.
This tentative duality is a strong/weak coupling duality in the sense that when 
$k\to \infty$, with $N_c$, $N_f$ fixed, the electric description becomes weakly 
coupled, whereas the magnetic description becomes strongly coupled 
\cite{Giveon:2008zn}. 

There are several aspects of this work that deserve further study. For instance,
the above theories contain a superpotential interaction $W\propto \tr X^{n+1}$ by an
operator that is classically irrelevant when $n>3$. We proposed that this is a
dangerously irrelevant operator. It would be interesting to verify this explicitly.

In this note we did not discuss extensively the case of two adjoint chiral superfields. 
As we mentioned in the main text, this case can be achieved by considering the 
general $n,n'$ brane setups of figs.\ \ref{CSQCD}, \ref{CSQCDmag}. What needs 
to be understood better is the CS theory that arises in this system at low energies.
Naively, this is a theory with fractional CS level. This is perfectly consistent
for $U(1)$ gauge groups, however, it is inconsistent with the quantization of 
the level for non-abelian gauge groups. It has been proposed that in this case
extra interactions need to be taken into account \cite{Bergman:1999na,Gaiotto:2008ak}.

We hope to return to some of these issues in a future publication.

\medskip
\section*{Acknowledgements}
\noindent
We acknowledge partial financial support by the INTAS grant, 03-51-6346, 
CNRS PICS \#~2530, 3059 and 3747 and by the EU under the contracts 
MEXT-CT-2003-509661, MRTN-CT-2004-503369 and MRTN-CT-2004-005104.

\newpage 

\end{document}